\documentclass[twocolumn,reprint,amsmath,amssymb,cha,aip,prb]{revtex4-1}

\usepackage{docs}
\usepackage{epsfig}% Include figure files
\usepackage{graphicx}% Include figure files
\usepackage{bm}%
\usepackage[colorlinks=true,linkcolor=blue]{hyperref}%

\begin{document}

\title{Interplay of electron correlation and covalency in FeTe$_{0.6}$Se$_{0.4}$}

\author{Ganesh Adhikary,$^1$ Deepnarayan Biswas,$^1$ Nishaina
Sahadev,$^1$ Swetarekha Ram,$^2$ V. Kanchana,$^2$ C. S. Yadav,$^1$
P. L. Paulose,$^1$ and Kalobaran Maiti$^1$\footnote{Corresponding
author: kbmaiti@tifr.res.in}}

\address{$^1$Department of Condensed Matter Physics and Materials
Science, Tata Institute of Fundamental Research, Homi
Bhabha Road, Colaba, Mumbai - 400 005, INDIA.\\
$^2$Department of Physics, Indian Institute of Technology Hyderabad,
Ordnance Factory Estate, Yeddumailaram, 502205, Andhra Pradesh,
India}

\begin{abstract}
We investigate the electronic structure of FeTe$_{0.6}$Se$_{0.4}$
employing high resolution photoemission spectroscopy and \emph{ab
initio} band structure calculations. Fe 2$p$ core level and the
valence band spectra exhibit signature of strong electron
correlation in the electronic structure. The electronic states near
the Fermi level reduces in intensity with the decrease in
temperature in conformity with the insulating transport observed
near 300 K. The observation of an insulator to metal transition
around 150 K in the transport properties may be related to the
spectral lineshape change in the vicinity of the Fermi level
observed in this study. The spectral features near Fermi level
exhibit significant $p$ orbital character due to the correlation
induced Fe $d$ spectral weight transfer. The experimental spectra
reveal dominant temperature dependence of the spectral functions
possessing large $p$-character. These results demonstrate
significant renormalization of the character of the conduction
electrons due to electron correlation and emphasizes the importance
of ligand states in the superconductivity of these materials.
\end{abstract}

\date{\today}


\maketitle

\section{Introduction}

Discovery of superconductivity led to an enormous growth of research
in both fundamental science and technology. While these materials
are extensively used in medical, scientific and engineering tools,
potential application in lossless power transmission, fast transport
system etc. has been the major driving force in the search of new
superconducting materials with high transition temperature. In this
respect, the discovery of superconductivity in Fe-based systems
\cite{Kamihara1,Kamihara2} renewed great attention due to the
complex interplay of magnetism and superconductivity.\cite{ganesh}
These materials are significantly different from copper oxide
superconductors.\cite{cuprate} It is believed that Fe 3$d$ states
derive the exotic properties in these systems unlike cuprates, where
the ligands play significant role.

Among Fe-based superconductors, Fe(TeSe) group of compounds are
believed to be the most correlated ones due to their large
`chalcogen height'\cite{Pnictogen height} (the height of the anions
from the Fe-plane)\cite{review}. These materials form in
anti-PbO-type crystal structure (space group
$P4/nmm$).\cite{structure} The parent compounds, FeTe exhibits a
spin density wave (SDW)-type antiferromagnetic transition at 65 K
\cite{FeTe_dos,FeTe_sdw} and FeSe is a superconductor below 8
K.\cite{FeSe1,FeSe2} Homovalent substitution of Te at Se-sites
introduces disorder in the system that is expected to reduce the
superconducting transition temperature, $T_c$.\cite{nandini} In
contrast, $T_c$ increases in this system with the maximum $T_c$ of
15 K for about 60\% of Te concentration,\cite{CSYadav1} where
disorder is expected to be the most prominent. Excess Fe
(Fe$_{1+y}$Te$_{1-x}$Se$_x$) in these materials exhibits magnetic
ordering with bicollinear commensurate/incommensurate structure that
survives even in the highest $T_c$ compound as a short range
order.\cite{baoPRL} Electronic structure studies show band narrowing
and large mass enhancement due to electron
correlation.\cite{ARPES-Nakayama,ARPES-Chen,ARPES-Tamai,Silke}

The normal phase (above $T_c$) of these compounds is complex
exhibiting deviation from metallic conductivity near 300 K similar
to that often observed in small gap semiconducting systems. Lowering
of temperature leads to a transition to metallic conductivity around
150 K and eventually superconductivity appears below 15 K.
Evidently, Fe(TeSe) exhibits plethora of complexity such as large
electron correlation, curious disorder effects, complex normal phase
etc. placing them in the pathway between other Fe-superconductors
and cuprates. Here, we studied the electronic structure of
FeTe$_{0.6}$Se$_{0.4}$ employing high resolution photoemission
spectroscopy and band structure calculations. Our results reveal
anomalous temperature evolution of the spectral function when the
probing photon energy is varied that could be explained in terms of
matrix element effect in photoemission and correlation induced
enhanced chalcogen $p$-contributions near Fermi level. The observed
change in electrical resistivity around 150 K has also been found to
influence the photoemission spectra.

\section{Experiment}

The single crystalline sample of FeTe$_{0.6}$Se$_{0.4}$
\cite{CSYadav2} was grown by flux method and characterized by
$x$-ray diffraction, Laue, M\"{o}ussbauer and tunneling electron
microscopic measurements establishing stoichiometric and homogeneous
composition of the sample with no trace of additional Fe in the
material. The photoemission measurements were carried out using a
R4000 WAL electron analyzer from Gammadata scienta, monochromatic Al
$K\alpha$ ($h\nu$ = 1486.6 eV), He {\scriptsize I} ($h\nu$ = 21.2
eV) and He {\scriptsize II} ($h\nu$ = 40.8 eV) photon sources and an
open cycle helium cryostat, LT-3M from Advanced Research Systems.
The sample was cleaved at a base pressure better than
4$\times$10$^{-11}$ torr at each temperature just before the
measurements. The energy resolutions were fixed to 2 meV, 5 meV and
350 meV for He {\scriptsize I}, He {\scriptsize II} and Al $K\alpha$
energies, respectively.

Band structure calculations were carried out using full-potential
linearized augmented plane wave method using WIEN2k
software.\cite{Wien2k} In local density approximations (LDA), local
uniform charge densities are considered to calculate the exchange
correlation functionals. To improve upon this approximation, the
exchange correlation is treated within the generalized gradient
approximation (GGA),\cite{gga} where the gradient terms of the
electron density are added in the exchange correlation functionals.
The GGA+$U$ method \cite{gga+u} ($U$ = on-site electron correlation
strength) was employed for $d$ electron interactions. The effective
on-site Coulomb repulsion strength, $U_{eff}$ = 0 - 7 eV ($U_{eff} =
U - J$; $J$ = Hund's exchange integral) was used to visualize the
correlation induced changes in the electronic structure.
2$\times$2$\times$2 supercell was used to generate the structure of
FeSe$_{0.5}$Te$_{0.5}$. A (25$\times$25$\times$7) $k$-point mesh
corresponding to 364 $k$ points in the irreducible part of the
Brillouin zone in the Monkhorst-Pack \cite{bz} scheme was used
during the self-consistent cycle.

\section{Results and Discussions}

\begin{figure}
% \vspace{-4ex}
\includegraphics [scale=0.4]{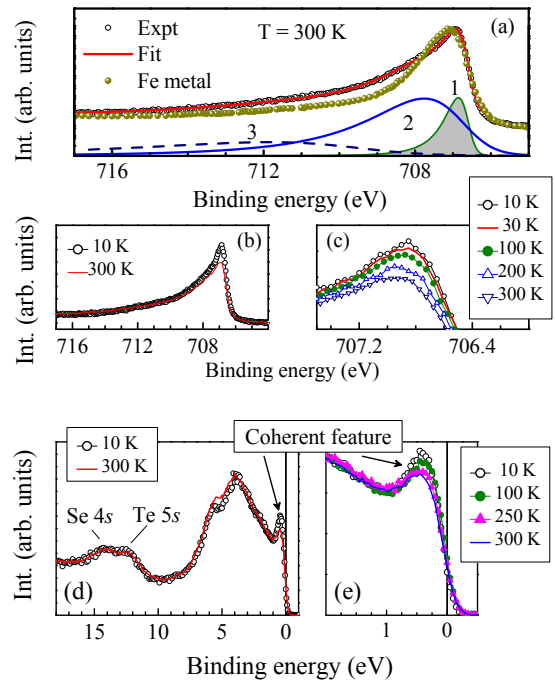}
 \vspace{-4ex}
\caption{(a) Fe 2$p$ spectra of Fe(TeSe) (open circles) and Fe metal
(closed circles - adapted from Ref. \cite{acker}). Lines show the
fit to the experimental spectrum. (b) Fe 2$p$ spectra at 300 K and
10 K. (c) Expanded view of the temperature evolution of Fe
2$p_{3/2}$ peak. (d) Valence band spectra at 300 K and 10 K. (e)
Expanded view near Fermi level region.}
% \vspace{-2ex}
\end{figure}

In Fig. 1(a), we show the Fe 2$p$ core level spectra exhibiting
interesting evolution with temperature and unusually large asymmetry
relative to that of Fe metal.\cite{acker} The low energy excitations
across the Fermi level in the photoemission final state lead to
asymmetry in the spectral lineshape of metallic systems. However,
the significantly large asymmetry in the present case requires
consideration of three asymmetric features as denoted by 1, 2 and 3
in Fig. 1(a). These features can be attributed to the well screened
final state (peak 1) and the other final states corresponding to
different electron-hole excitations as observed in earlier studies
of other related systems.\cite{takahashi} While the peak 1 is
distinctly visible due to better energy resolution in this study,
its intensity relative to the higher binding energy features is
weaker than that in Fe indicating poorer itineracy of the conduction
electrons in FeTe$_{0.6}$Se$_{0.4}$. Interestingly, the intensity of
the peak 1 increases gradually with the decrease in temperature (see
Figs. 1(b) and 1(c)).

Earlier theoretical studies\cite{RMP-Gabi} based on dynamical mean
field theory showed that in a strongly correlated system, the
valence band spectra exhibit two features. One feature appears at
the Fermi level called coherent feature, which represents the
itinerant electrons. The second feature, termed as incoherent
feature or lower Hubbard band, corresponds to the correlation
induced localized states and appears at higher binding energy.
Decrease in temperature leads to an enhancement of the coherent
feature intensity with a consequent decrease in the incoherent
feature intensity. The valence band spectra in Figs. 1(d) and 1(e)
exhibit scenario similar to the above theoretical results; the
intensity near the Fermi level gradually increases with the decrease
in temperature relative to the intensities at higher binding
energies. This behavior of the spectral function is a manifestation
of the correlation induced effect and also suggests that the
intensity of itinerant electrons increases with the decrease in
temperature. The enhancement of the well screened feature in the Fe
2$p$ core level spectra (peak 1) with the decrease in temperature is
consistent with this scenario.

\begin{figure}
% \vspace{-4ex}
\includegraphics [scale=0.4]{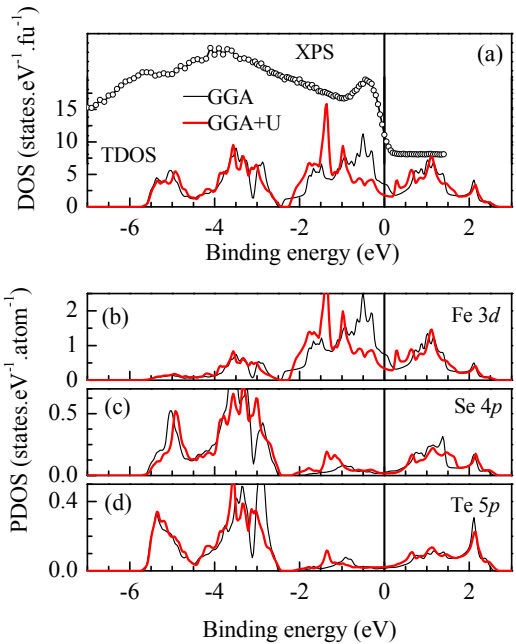}
 \vspace{-14ex}
\caption{(a) Calculated total density of states for uncorrelated and
correlated system ($U$ = 4.0 eV) and the experimental spectrum. (b)
Fe 3$d$, (c) Se 4$p$ and (d) Te 5$p$ PDOS from GGA (thin line) and
GGA+$U$ (thick line) calculations with $U$ = 4.0 eV.}
% \vspace{-4ex}
\end{figure}

In order to investigate the correlation induced effects further, the
calculated electronic density of states are compared with the
experimental spectra in Fig. 2. Finite electron correlation leads to
a spectral weight transfer from $\epsilon_F$ to higher binding
energies leading to an enhancement of intensity around 2 eV. Since
the transferred spectral weight appears in the energy range similar
to the energy range of bonding bands, distinct identification of the
incoherent feature is non-trivial in this system. We observe that
the calculated spectral functions in the whole valence band energy
range with finite $U$ exhibit better description of the experimental
spectrum. For example, the calculated spectrum for $U~=~$4~eV is
shown in the figure providing a good description of the experimental
features.

It is well established that covalency plays significant role in the
electronic structure of various (3$d$, 4$d$ amd 5$d$) transition
metal compounds\cite{covalency}. Energy bandwidth depends of the
degree of hybridization with the neighboring sites. Therefore,
electron correlation would affect the electronic states with
different orbital character differently depending on their degree of
itineracy in the uncorrelated system. This is evident in Fig. 2
exhibiting significant transfer of the Fe 3$d$ partial density of
states (PDOS) to higher binding energies. Subsequently, the Se
4$p$~/~Te 5$p$ contributions increase near $\epsilon_F$. These
results suggests strong renormalization of Fe $d$ - chalcogen $p$
covalency in this system.

\begin{figure}
% \vspace{-4ex}
\includegraphics [scale=0.4]{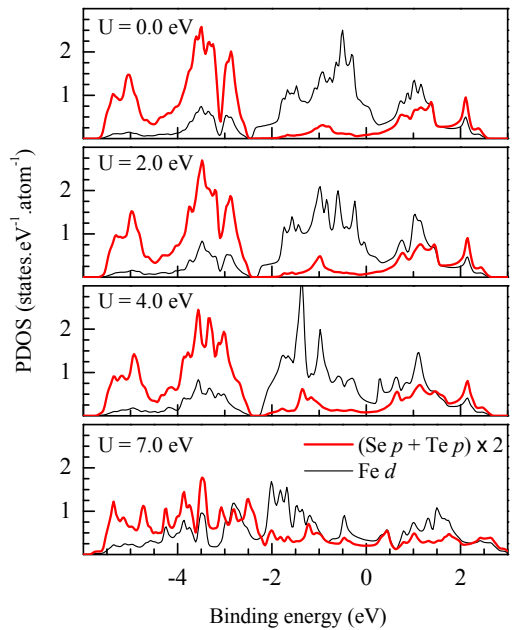}
 \vspace{-14ex}
\caption{Calculated partial density of states for Fe 3$d$ and
rescaled $p$-states [(Te 5$p$ + Se 4$p$)~$\times$~2] for different
values of $U$. The contribution of $p$-states near the Fermi level
gradually increases with respect to the Fe 3$d$ contributions with
the increase in $U$.}
% \vspace{-4ex}
\end{figure}

The change in $p$ and $d$ contributions with the increase in $U$ is
demonstrated in Fig. 3. The electronic states near $\epsilon_F$ is
primarily dominated by the Fe 3$d$ contributions for $U$ = 0.0 eV.
With the increase in $U$, the Fe 3$d$ partial density of states
gradually shifts towards higher binding energies with substantial
increase in intensity around 2 eV. Subsequently, the chalcogen $p$
contributions near $\epsilon_F$ gradually increases making the
covalent mixing much stronger compared to the uncorrelated case.

\begin{figure}
% \vspace{-4ex}
\includegraphics [scale=0.4]{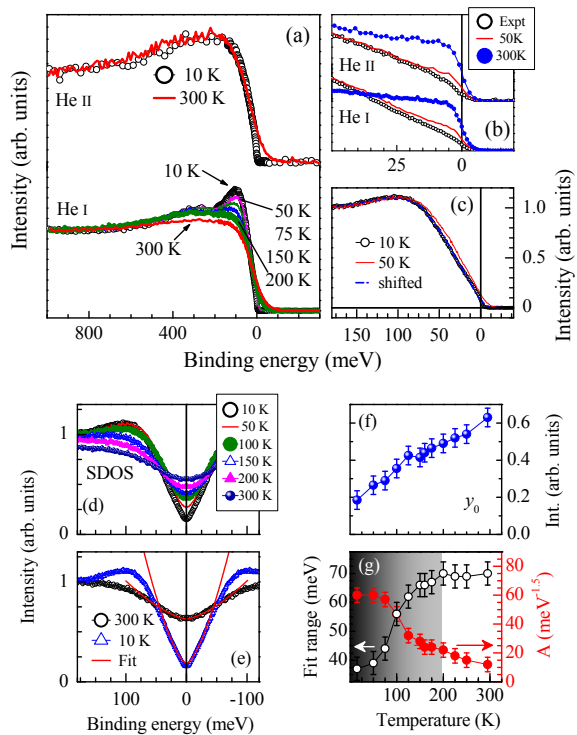}
 \vspace{-4ex}
\caption{(a) He {\scriptsize I} and He {\scriptsize II} spectra. (b)
10 K data (open circles) is compared with extracted 10 K spectra
from 300 K (solid circles) and 50 K (line) data. (c) 50 K data
(solid line) and shifted 50 K data (dashed line) are compared with
10 K data (open circles) suggesting a gap of 5 meV. (d) SDOS
obtained by symmetrization at different temperature. (e) Fit for 300
K and 10 K SDOS. The fitting parameters, (f) $y_0$, and (g) $A$ \&
fit range.}
% \vspace{-6ex}
\end{figure}

The high resolution spectra exhibit curious evolution with
temperature. The He {\scriptsize II} spectra in Fig. 4(a) exhibit a
broad feature near $\epsilon_F$ with no distinct change in lineshape
with temperature apart from a sharp fall in intensity at
$\epsilon_F$ due to the superconducting gap below
$T_c$.\cite{ramarao} In contrast, the He {\scriptsize I} spectra
exhibit two distinct features at 0.1 and 0.3 eV. The intensities of
these peaks increase gradually with the decrease in temperature.
Since the spectra were collected using transmission mode with a
large acceptance angle of $\pm$15$^o$, the experimental spectra will
be a representation of the total density of states. Thus, the change
in spectral features might have a origin different from momentum
dependence. The photoemission cross section for Fe 3$d$ states is
4.833 and 8.751 at 21.2 eV and 40.8 eV photon energies,
respectively. While that of As 4$p$ states is 3.858 and 0.2949,
respectively. Thus, the relative photoemission
cross-section\cite{yeh} of As 4$p$ to that of Fe 3$d$ states
enhances from 0.034 at 40.8 eV to 0.8 at 21.2 eV. Since the valence
states are well described by the linear combination of atomic
orbitals, one can conclude that the contribution of $p$ orbital
character in the conduction electronic states is significant and
exhibit higher degree of sensitivity to the change in temperature.

The spectral density of states, SDOS at different temperatures are
estimated by the division of the high resolution spectra by the
energy resolution broadened Fermi-Dirac function. The representation
of the 10 K spectra is a convolution of the above SDOS by the
resolution broadened Fermi function of 10 K. Thus obtained 10 K
spectra based on 300 K and 50 K spectra are compared with the raw
data at 10 K in Fig. 4(b). The 300 K spectra exhibit large intensity
at $\epsilon_F$. The intensity gradually decreases with temperature
as also seen in the SDOS obtained by symmetrization ($I(\epsilon) =
I_0(\epsilon - \epsilon_F) + I_0(\epsilon_F - \epsilon)$) in Fig.
4(d). Such decrease in carrier concentration at $\epsilon_F$ with
the decrease in temperature independent of the analysis procedure
adopted might be the origin for the absence of metallic resistivity
behavior near 300 K although the density of states indicate metallic
phase. The photoemission spectra, however, exhibit monotonic
decrease in intensity at $\epsilon_F$ with the decrease in
temperature that cannot capture the transition to metallic
temperature dependence of resistivity below 150 K. There are
prediction of the persistence of short range antiferromagnetic order
in this composition \cite{baoPRL} that might introduce such anomaly
in resistivity.

In Fig. 4(c), the raw data at 50 K and 10 K are observed to be
shifted due to the formation of the superconducting gap below $T_c$.
A shift of 5 meV of the 50 K data towards higher binding energy
reproduces the rising part of the spectrum at 10 K excellently well
suggesting the superconducting gap of about 5 meV. The lineshape of
the spectral functions is investigated by fitting the raw data with
a polynomial; the best fit is found for the relation, $I(\epsilon) =
y_0 + A\times |\epsilon - \epsilon_F|^{1.5}$ as demonstrated in Fig.
4(e). The fitting parameters are shown in Figs. 4(f) and 4(g). The
observed energy exponent of 1.5 indicates importance of magnetic
fluctuations in the ground state.\cite{bairo3} While the fitting
range starts limiting itself in the low energy regime below 150 K,
coefficient, $A$ for the spin fluctuation term starts becoming
larger at the same temperature, where the transition to metallic
phase occurs indicating their importance in the low temperature
properties.

Effect of disorder has extensively been studied theoretically and
experimentally.\cite{tvr} Altshuler and Aronov showed that charge
disorder in correlated systems exhibits $|\epsilon -
\epsilon_F|^{0.5}$ dependence of the spectral lineshape near Fermi
level and the spectral intensity at $\epsilon_F$ is proportional to
the square root of temperature.\cite{altshuler,DD-disorder} In the
present case, SDOS at $\epsilon_F$, represented by $y_0$ in Fig.
4(f) does not follow the typical square root dependence on
temperature in conformity with the above observation indicating
importance of magnetic interactions in these systems.

\section{Conclusions}

In summary, we studied the electronic structure of Fe-based
superconductor, FeTe$_{0.6}$Se$_{0.4}$ employing high resolution
photoemission spectroscopy and high quality single crystalline
sample. The core level and valence band spectra exhibit signature of
electron correlation induced features. The intensity at the Fermi
level reduces gradually with the decrease in temperature suggesting
decrease in conductivity as manifested by insulating temperature
dependence of electronic conduction in transport measurements. The
transition from insulating to metallic conductivity could be related
to the change in spectral lineshape near the Fermi level.

Electron correlation renormalizes the covalency significantly
leading to a significant chalcogen $p$ orbital contributions near
the Fermi level. Interestingly, the electronic spectra possessing
large $p$ character exhibit stronger sensitivity to the change in
temperature. The spectral lineshape and its temperature evolution
indicate importance of magnetic fluctuations in the electronic
properties. Evidently, ligands play an important role in high
temperature superconductivity as also found in cuprates.

\section{Acknowledgements}

The authors N. S. and K. M. acknowledge financial support from the
Dept. of Science and Technology, Govt. of India under the
Swarnajayanti fellowship programme.

%\section*{References}

\end{document}